\documentclass[12pt]{iopart}
\usepackage{iopams}
\usepackage{graphicx}
\begin{document}

\title[K.K. Yadav et al.]{Search for  TeV $\gamma$ -rays from H1426+428 during 2004-07 with  the TACTIC telescope}

\author{ K.K. Yadav$^1$, R.C. Rannot$^1$, P. Chandra$^1$, A.K. Tickoo$^1$, S. Thoudam$^1$, 
K. Venugopal$^1$, N. Bhatt$^1$, S. Bhattacharyya$^1$, K. Chanchalani$^1$, V.K. Dhar$^1$, S.V. Godambe$^2$, H.C. Goyal$^1$, M. Kothari$^1$, S. Kotwal$^1$, M.K. Koul$^1$, R. Koul$^1$,
S. Sahaynathan$^1$, M. Sharma$^1$}

\address {$^1$   Astrophysical Sciences Division, Bhabha Atomic Research Centre, \\
Mumbai - 400 085, India.}
\address {$^2$  Physics Department, University of Utah, Salt Lake City, UT 84112, USA.}
\ead{kkyadav@barc.gov.in}

%-----------------------------------------------------------------------------------------------%
\begin{abstract}
The  BL Lac object H1426+428 ($z\equiv 0.129$) is an established source of TeV $\gamma$-rays and detections of these photons from this object  also have important implications for  estimating the Extragalactic Background Light (EBL) in addition to the understanding of the  particle acceleration and $\gamma$-ray production mechanisms in the AGN jets. We have observed this source for about 244h in 2004, 2006 and 2007 with the TACTIC $\gamma$-ray telescope located at Mt. Abu, India. Detailed analysis of these data do not indicate the presence of any  statistically significant TeV $\gamma$-ray signal from the source direction. Accordingly, we have    placed an upper limit of $\leq1.18\times10^{-12}$ $photons$ $cm^{-2}$ $s^{-1}$ on the integrated  $\gamma$-ray flux at 3$\sigma$ significance level.
\end{abstract}

%-----------------------------------------------------------------------------------------

%--------------------------------------------------------------------------------------------------------
\section{Introduction}
BL Lacertae (BL Lac) objects are a subclass of blazars that are notable for their lack of prominent emission lines. The Spectral  Energy  Distributions (SED)  of these objects extend from radio to $\gamma$-ray energies with the double humped  shape  feature in a $\nu F_\nu$ representation \cite{bed97,Ulrich97,UrryPadovani95}. The broadband  SED of the BL Lac objects  identified in X-ray surveys are significantly different compared to those identified in radio surveys and have been classified  into high frequency peaked BL Lac objects (HBL) and low frequency peaked BL Lac objects (LBL), depending on their X-ray to radio flux densities \cite{Padovani95}.
The first hump in SED of these objects is expected at X-ray energies and generally accepted as originating from synchrotron radiation of relativistic electrons in the magnetic field around the object. The second peak is expected at $\gamma$-ray energies and in the framework of leptonic models \cite{Bloom96,Konigl81,Maraschi92,Sikora01,Blandford95,Dermer97,Ghisellini96} is thought to be stemming from inverse Compton scattering of low energy photons to $\gamma$-ray energies by the same population of relativistic electrons which  produces the synchrotron radiation.
Whereas, the hadron based  models \cite{Bednarek93,DarLaor97,Aharonian00,MP01} attribute the second hump and X-ray emission to some extent  to processes involving  protons which are also accelerated along with the leptons  in the  relativistic  jets.

These objects  are extremely variable and are characterized by low emission quiescent states with occasional flaring when the flux can increase by several orders of magnitude. For example, the HESS group  detected the  activity of the  object  PKS2155 -304 in July 2006 when its emission reached a flux level more than two orders of magnitude higher than its quiescent flux \cite{Benbow07}.  

Very High Energy (VHE) $\gamma$-ray photons are believed to be  absorbed by their interactions with the extragalactic background light(EBL) leading to an energy dependent horizon for viewing of VHE $\gamma$-ray sources. As a result of this absorption the energy spectrum of such sources may show a cut-off feature or steepening of the spectrum, but at the same time such features can also be attributed to the source itself. The interpretation of such features can be used as a probe for EBL measurements \cite{Stecker92}. Interestingly, the recent detections of the AGN 3C 279 (z = 0.536) by the MAGIC group \cite{magic3c279} and,  H2356-309(z = 0.165) and  1ES 1101-232 (z = 0.186) by the HESS group \cite{hess_ebl} have incited a lot of interest in the understanding of the $\gamma$- ray horizon at very high energies. 
\par

H1426+428 was discovered in the 2-6 keV energy band by HEAO 1 \cite{Wood84} and was classified as a BL Lac object in 1989 \cite{Remillard89}. In an observation campaign performed by BeppoSAX in 1998-99, this source was identified as an extreme HBL because of its high synchrotron peak frequency, therefore a potential source of VHE $\gamma$ -rays \cite{Costamante01}.
\par
The  Whipple group reported the first VHE detection of this source in 2002, using the data recorded during   1999 -2001, at a statistical significance of about  $6\sigma$ \cite{Horan02}.   
Subsequently it was confirmed by the CAT\cite{Djannati02} and HEGRA groups \cite{Aharonian02,Aharonian03} at  statistical significance of $5.2\sigma$ and $7.5\sigma$ respectively. 
The CELESTE and STACEE groups, both with non-imaging detectors have also observed the source and  reported  upper limits  during  the periods 2002- 2004   \cite{david06} and  2001- 2002 \cite{stacee} respectively.
However, the Crimean group has reported the detection of VHE $\gamma$-rays from this source using their GT-48 system in April 2004 with  a statistical significance of 5.8$\sigma$\cite{GT-48}. 
More recently, the VERITAS group  observed this source for 12.5 h using their  stereoscopic telescope  system at Mt. Hopkins. Preliminary results obtained  by them indicate a marginal excess of VHE photons  at 3.2 $\sigma$ statistical level  and they have placed an upper limit of about 3$\%$ of the Crab Nebula flux \cite{kraw07}.   
 In this paper, we report the results of our  observations of this  source which were made during the period 2004-07  using the TACTIC $\gamma-$ray telescope. In addition, we have also compared the TACTIC light curves with those of RXTE/ASM X-ray in the energy range 2-10 keV.  
%-------------------------------------------------------------------------------------------------------------
\section{ TACTIC telescope}
The TACTIC (TeV Atmospheric Cherenkov Telescope with Imaging Camera) $\gamma$-ray telescope has been set up at Mt. Abu ( 24.6$^\circ$ N, 72.7$^\circ$ E, 1300m asl), India for studying the emission of TeV $\gamma$-rays from celestial sources. The telescope deploys a F/1 type tracking light collector of $\sim$9.5 m$^2$ area, made up of 34 x 0.6 m diameter, front-coated spherical glass facets which have been prealigned to produce an on-axis spot of $\sim$ 0.3$^\circ$ diameter at the focal plane. The telescope uses a 349-pixel photomultiplier tube (ETL 9083UVB) -based imaging camera  with a uniform pixel resolution of $\sim$0.3$^\circ$ and a field-of-view of $\sim$6$^\circ$x6$^\circ$ to record images of atmospheric Cherenkov events. The present data have been collected with the inner 225 pixels where the innermost 121 pixels are used for generating the event trigger. The trigger is based on the 3NCT (Nearest Neighbour Non-Collinear Triplets) configuration  for 2004 and 2007 data while 2006 data were collected with NNP (Nearest Neighbour Pairs) trigger logic \cite{Kaul03}. The safe operation of the photomultipliers (PMT) with anode current $\leq$ 3 $\mu$A  has been ensured by implementing a gain control algorithm \cite{Bhatt01}. The data acquisition and control system of the telescope \cite{Yadav04} have been designed around a network of PCs running the QNX (version 4.25) real-time operating system. The single pixel threshold was set to $\geq$ 8/25 pe  for 3NCT/NNP trigger logic. The triggered events are digitized by CAMAC based 12-bit Charge to Digital Converters (CDC) which have a full scale range of 600 pC. The sensitivity of this telescope  is  such that it detects VHE $\gamma$-rays from the Crab Nebula direction at a statistical significance of  5$\sigma$  in 25h above $\sim$1.5 TeV.  The details of the telescope can be found in \cite{Koul07}. 
%------------------------------------------------------------------------------------------------------------
\section{Observations and data analysis}
H1426+428 was observed  for 244h from 2004 to 2007 with the TACTIC $\gamma$-ray telescope. In order to maximize the on-source observation time and to increase  the possibility of recording  flaring activity from the source, we have made observations  in the continuous source tracking  mode. This mode of observation is also known as a discovery mode wherein  the   systematic errors are relatively more as compared to the  ON-OFF mode  of data taking (where the background is estimated from separate observations). In the observation mode  used for the ongoing  studies, the source  was always placed at the centre of the camera while making observations on it. The zenith angle range covered  during these on-source observations  was from 18$^\circ$ to  45$^\circ$. Details of these observations have been given in  Table \ref{tab:obs}.
\par
While analysing, the data were subdivided into 3 spells each  corresponding  to  the observations made in a particular year.  Standard data quality checks, e.g. compatibility of the prompt and chance coincidence rates with Poissonian statistics, have been performed in order to evaluate the system behavior and the general quality of the recorded data. As a result of this exercise,  77.9 h of data were rejected and the rest of the data were used for detailed analysis.
\begin{table}[h]
\caption{ Details of TACTIC Observations on H1426+428.}
\centering
\begin{tabular}{|c|c|c|c|c|c|c|}
\hline 
Spell &Year & Months  & Observation & Total data & Data selected \\
     &      &         & Dates  & (h) & (h )\\
%Spell &Year & Months  & Observation dates & Observation time(h) \\
\hline
I & 2004 & Mar. & 22-26, 28,29,31 & 71.0 & 44.8\\
  &      & Apr. & 12-15, 18, 20-22, 24-26 & &\\ 
\hline
II & 2006 & Mar. & 26-31 &  &  \\
   &      & Apr. & 1, 2, 4, 21, 23-30 &34.9 & 18.4\\
   &      & May  & 01, 02, 19, 21, 23-24, 29-30 & & \\
\hline
III & 2007 & Mar. & 16-27 & & \\
    &      & Apr. & 12-15, 18-20, 22-26 &137.7 & 102.5\\
    &      & May  & 06-15, 17-19, 21-22  & & \\
    &      & Jun. & 07-11  & & \\
\hline
\end{tabular} 
\label{tab:obs}
\end{table}   
%------------------------------------------------------------------------------------- 

While performing detailed analysis of the data,  firstly all the data were  subjected to pedestal removal by using the standard two-level image 'cleaning' procedure \cite{Koul07} with picture and boundary thresholds of 6.75$\sigma$ and 3.25$\sigma$ respectively. The digital counts of each pixel in the clean images were then corrected for inter-pixel gain variation. The image cleaning threshold levels were first optimized on the Crab Nebula data \cite{Godambe08} and then applied to the data presented in this paper. The clean Cherenkov images were characterized by calculating their standard image parameters like LENGTH, WIDTH, DISTANCE, ALPHA, SIZE and FRAC2
\cite{Hillas85,Konopelko96}. The standard Extended Supercuts \cite{Mohanty98} procedure was then used to separate $\gamma$-ray like images from those  due to  the huge background of cosmic rays. The $\gamma$-ray selection criteria based on the imaging  cuts, given in    Table \ref{tab:cuts} used in the analysis have been obtained on the basis of   Monte Carlo simulations carried out for  the TACTIC telescope.

\begin{table}[h]
\caption{ Extended Supercuts  selection  criteria used for analyzing the TACTIC data.}
\centering
\begin{tabular}{|c|c|}
\hline 
Parameter  & Cut Values\\
\hline
LENGTH (L) & $0.11^\circ\leq L \leq(0.235+0.0265 \times \ln S)^\circ$\\
\hline
WIDTH  (W) & $0.065^\circ \leq W \leq (0.085+0.012 \times \ln S)^\circ$\\
\hline
DISTANCE (D) & $0.5^\circ\leq D \leq 1.27^\circ$ \\
\hline
SIZE (S)  & $S \geq 350  d.c$; (6.5 digital counts $\equiv$ 1.0 pe )\\
\hline
ALPHA ($\alpha$) &  $\alpha \leq 18^\circ$\\
\hline
FRAC2 (F2) &  $F2 \geq 0.38$ \\
\hline
\end{tabular} 
\label{tab:cuts}
\end{table}  
%--------------------------------------------------------------------------------------

\par                
The next step of the data analysis deals with estimating  the number of $\gamma$- ray events from the source direction in the presence of overwhelming background of cosmic ray events. This is done by plotting the frequency distribution of the 'Alpha' parameter (
defined as the angle between the major axis of the image and the line joining the image centroid with the source position
) of shape, size and distance selected events. The $\gamma$-rays  from a point source have smaller values of the 'Alpha' parameter, whereas the cosmic ray events, because of their isotropic nature, are uniformly distributed in all 'Alpha' values ranging  from 0$^\circ$  to 90$^\circ$. 
For the TACTIC system we find that the 'Alpha' range for $\gamma$-ray events is found to be from  $0^\circ\  to \leq18^\circ$.

The flat distribution of 'Alpha' in the range $27^\circ\leq\alpha\leq81^\circ$ has been used to determine the background events present in the signal region \cite{Yadav07}. The number of $\gamma$-ray events is then calculated by subtracting the expected number of background events (calculated on the basis of the background region) from the  $\gamma$-ray domain events and the significance of the excess events has been finally calculated by using the maximum likelihood ratio method of Li $\&$ Ma \cite{Li83}.
\par

\begin{figure}[h]
\centering
\includegraphics*[width=1.0\textwidth, angle=0,clip]{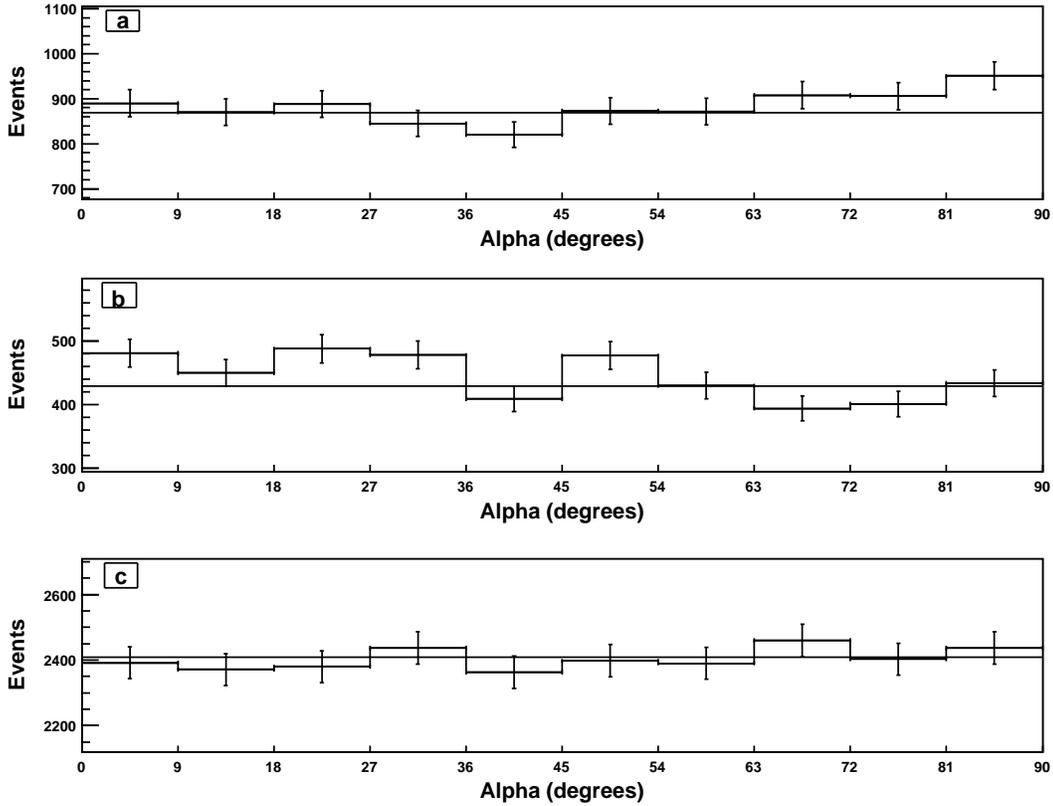} 
\caption {(a) Distribution of image parameter 'Alpha' after applying the cuts given in Table \ref{tab:cuts} for 44.8h 18.4h and 102.5h of data taken on H1426+428 during (a) 2004, (b) 2006 and (c) 2007 respectively. The horizontal lines in the figure indicate the expected background in the $\gamma$-domain obtained by using the background region ($27^\circ\leq\alpha\leq81^\circ$). The error bars shown are for statistical errors only.} \label{fig:alpha467}
\end{figure}

\subsection{TACTIC data analysis results}

\begin{figure}[h]
\centering

\includegraphics*[width=0.9\textwidth,angle=0,clip]{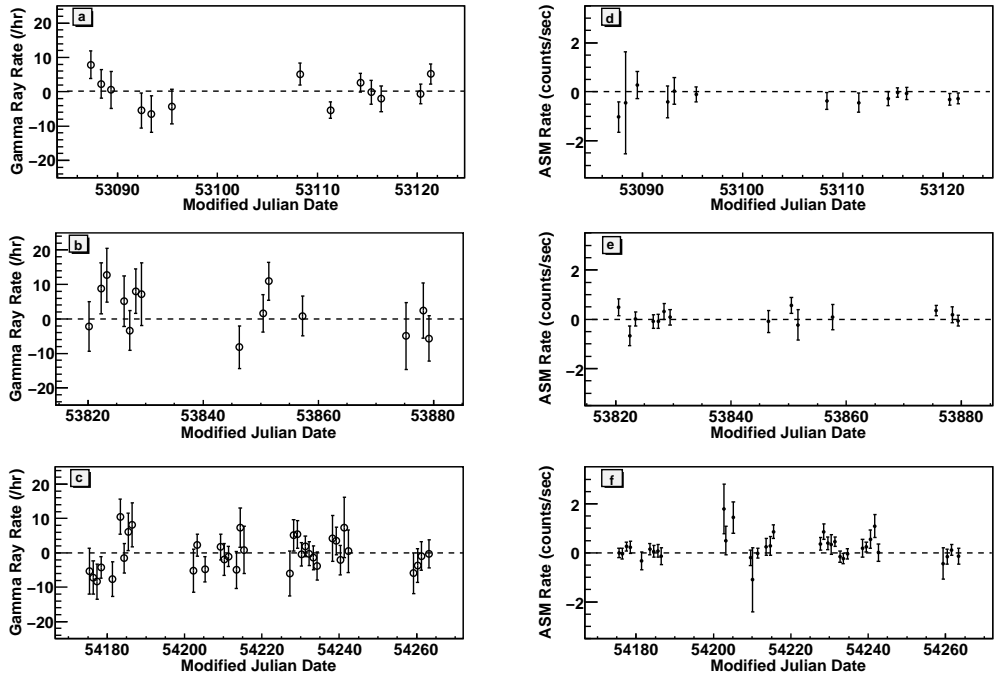}

\caption {TACTIC light curves for (a) 2004, (b) 2006 and (c) 2007 observations  and  corresponding ASM  light curves are shown in  (d), (e) and (f)  respectively. } \label{fig:lc}
\end{figure}

When all the  data recorded during the year  2004 are analysed together, the corresponding results obtained are shown in Figure \ref{fig:alpha467}a, wherein the histogram  of the 'Alpha' parameter has been plotted after having applied shape and orientation related imaging cuts given in Table \ref{tab:cuts}.  
As is clear from this figure, the  distribution  is almost flat and the number of gamma-ray like events within the $\gamma$- ray domain of the distribution are 19$\pm$48, thereby indicating that  the source was possibly  in a low TeV emission state (below TACTIC sensitivity level) during the period of these observations.
Further   the same  data were also analysed on a nightly basis to explore the possibility of  very strong episodic TeV emissions. The corresponding results obtained are depicted in Figure \ref{fig:lc}a,  which shows the day-to-day variations of the $\gamma$-ray rate ($\gamma$-rays/hour) for 2004  observations. This light curve is characterised with  a reduced $\chi^2$ value of  19.48/12 ( here and afterwards the denominator represents  the value for the degrees of freedom ) with respect to the zero degree polynomial fitted  constant value   of 0.28$\pm$0.97 photon events, with corresponding probability of  0.11  which is consistent with the no- variability hypothesis. 
The magnitude of an excess or deficit recorded on different nights is within $\pm$ 2 $\sigma$  level for 2004 observations and hence indicates the absence of a statistically significant episodic TeV gamma-ray signal from the H1426+428 direction.

\begin{figure}[h]
\centering     
\includegraphics*[width=0.9\textwidth,angle=0,clip]{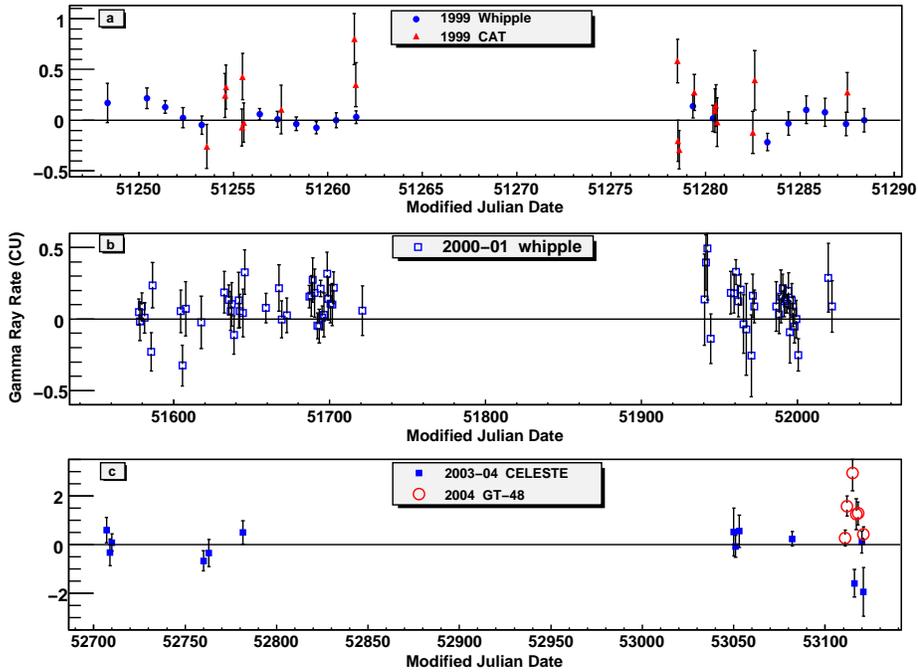}
\caption { H1426+428 TeV light curves   obtained by various groups, namely  Whipple\cite{Horan02}, CAT\cite{Djannati02}, CELESTE\cite{david06} and GT-48\cite{GT-48}.}  \label{fig:tev}
\end{figure}

In addition, the available TeV light curves reported  by different groups upto the year 2004, are  also depicted  in Figure \ref{fig:tev} starting from the Whipple 1999 observations. As is clear from this figure that the  GT-48 telescope has recorded the source at more than  200$\%$ of the GT-48 detected Crab level on April 19, 2004 (MJD 53115) at 3.9 $\sigma$ level\cite{GT-48,gt}. It may be noted here that  the TACTIC 2004  results discussed above on this source are in  agreement with those obtained with the CELESTE system \cite{david06} around 100GeV during the same period. However,  our results are in conflict with those  of GT-48 telescope  above 1TeV during the period from April 15 to April 25, 2004, wherein  a  $\gamma$ -ray signal at 5.8$\sigma$  level was reported\cite{GT-48}.

We have followed the same data analysis methodology while analysing the data recorded during the years 2006 and 2007 and  the corresponding 'Alpha'  histograms are shown in Figures  \ref{fig:alpha467}b and  \ref{fig:alpha467}c respectively. As is clear from these figures the  distributions  are almost flat and the number of gamma-ray like events within the $\gamma$- ray domain of the distributions are 68$\pm$35 and -54$\pm$80   
for the 2006 and 2007 observations  
respectively, thereby indicating that  the source was possibly  in a low TeV emission state (below TACTIC sensitivity level) during the period of these observations  as well. 
\begin{figure}[h]
\centering
\includegraphics*[width=1.0\textwidth,height= 0.5\textwidth,angle=0,clip]{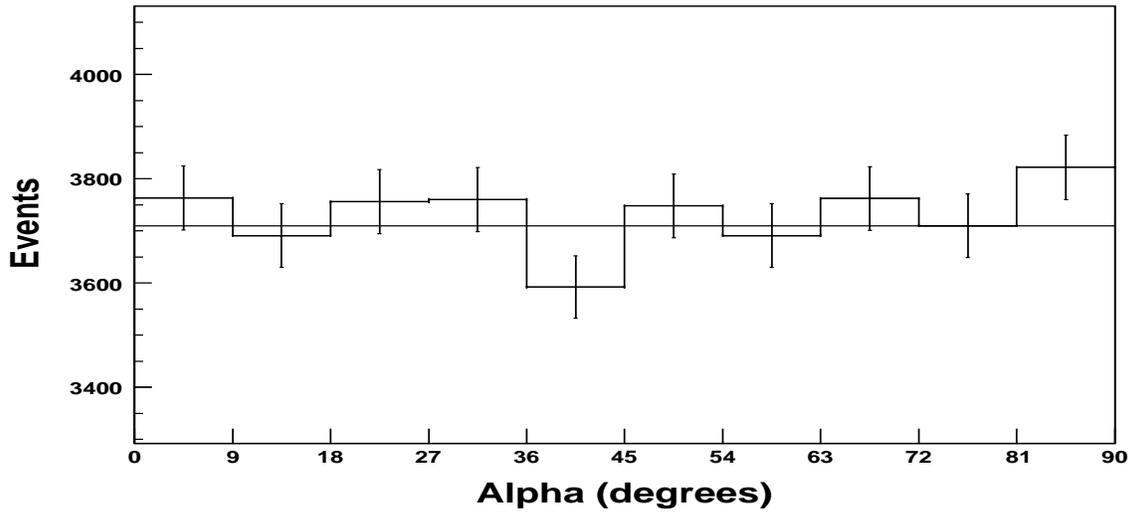}
\caption {Distribution of image parameter 'Alpha' after applying cuts given in Table \ref{tab:cuts} for the entire 165.7h of data on H1426+428. The horizontal line in the figure indicates the expected background in the $\gamma$-domain obtained by using the background region ($27^\circ\leq\alpha\leq81^\circ$). The error bars shown are for statistical errors only.} \label{fig:alpha_total}
\end{figure}
Further these data were also analysed on a nightly basis and  the corresponding results obtained are depicted in Figures \ref{fig:lc}b and \ref{fig:lc}c respectively, again showing  the day-to-day variations of the  hourly $\gamma$-ray rate with time.   Light curves are characterised with the reduced $\chi^2$ values of  12.66 /13  and 31.31 /34  respectively, with respect to the zero degree polynomial fitted  constant value of 2.22 $\pm$1.79 and -0.49$\pm$0.74  photon events with corresponding probabilities of  0.47 and 0.69 respectively which are consistent with the no- variability hypothesis. 
The magnitude of an excess or deficit recorded on different nights is also within $\pm$ 2 $\sigma$  level for 2006 and 2007  observations thereby again  indicating the absence of a statistically significant episodic TeV gamma-ray signal during these observations as well.
\par

In Figure \ref{fig:alpha_total}, we have shown the 'Alpha' histogram obtained when the data for all the years were combined together which is also consistent with the flat distribution. There is thus no evidence for the presence of a  statistically significant $\gamma$-ray signal from the source direction during the period of our observations. We have  placed an   upper limit of $\leq1.18\times10^{-12}$ $photons$ $cm^{-2}$ $s^{-1}$  at 3$\sigma$ level (13.1 $\%$ of the TACTIC detected Crab Nebula integrated $\gamma$-ray flux) on the integrated TeV $\gamma$-ray flux from the source direction, using the method of Helene \cite{Helene83}. In Table \ref{tab:results}, we have compiled all results obtained on a yearly spell basis as well as for the case  when  all the data are considered together. 

%During the detection of the source, the CAT group has observed the maximum flux of 0.8 Crabin 1999 while Whipple light curve shows the emission up to $\sim$0.3 Crab. The light curve obtained by GT-48 telescope during 2004 observations shows the the emission at $\sim$3 Crab level, which is the highest flux ever recorded from the source.

\par
\subsection{ASM X-ray light curves}
%In order to compare the TACTIC  light curves  with those of  RXTE/ASM X-ray  observations  in the energy range of 2-10 keV, we have used the daily average count rate of ASM from its archived data \cite{ASM} to obtain the light curves for the contemporary periods. 
In order to examine any correlation between VHE $\gamma$-ray and X-ray (2-10 keV) source emissions, the  Whipple group who discovered this source has done correlation studies for this source and no evidence for significant correlation was found, rather their 2001 data show some evidence for an anticorrelation between the average monthly rates\cite{Horan02}.
In the present work we also compare the TACTIC  light curves  with those of RXTE/ASM X-ray observations  in the energy range of 2-10 keV. For this we have used the daily average count rate of ASM from its archived data\cite{ASM} to obtain the light curves for the contemporary periods. These light curves are  shown in Figures  \ref{fig:lc}d,  \ref{fig:lc}e and  \ref{fig:lc}f for the corresponding TACTIC observation periods of 2004, 2006 and 2007 respectively. ASM light curves are characterised with  the reduced $\chi^2$ values of  5.27/12, 11.03/13 and 43.78/34  with respect to the zero degree polynomial fitted constant values of -0.2$\pm$0.088, 0.095$\pm$0.082,  and  0.152$\pm$0.046   counts  and corresponding probabilities of 0.96, 0.60 and 0.17 respectively, thereby  indicating that these light curves are  consistent with the constant flux hypothesis and no variability is observed  in a time scale of a day or more in the RXTE/ASM energy band also. We have not done  any correlation studies between the TACTIC and RXTE/ASM observations, mainly because of non detections of the source in the present work.

\begin{table}[h]
\caption{Yearly spell wise and consolidated data analysis results obtained on  H1426+428 with the TACTIC telescope.}
\centering
\begin{tabular}{|c|c|c|c|c|c|c|}
\hline 
Year & Obs. time & Excess/Deficit of events   &  Significance  & Upper Limit at 3$\sigma$ level     \\ 
&  ( h )  & from source direction       &  ( $\sigma$ )  & photons cm$^{-2}$ s$^{-1}$                \\
\hline
 2004    &  44.8  & 19$\pm$48    & 0.39       & $\leq2.13\times 10^{-12}$          \\
\hline 
 2006  &   18.4   & 68$\pm$35  & 1.95            & $\leq3.72\times10^{-12} $         \\
\hline
 2007  &  102.5   & -54$\pm$80   & -0.68          &$\leq1.54\times10^{-12}$       \\
\hline
 Total  &  165.7   & 33$\pm$100   & 0.33         &$ \leq1.18\times10^{-12} $          \\
\hline
\end{tabular}
\label{tab:results}
\end{table}

\section{Discussion and conclusions}    

We have observed the BL Lac object H1426+428  in  
the VHE gamma-ray energy range   with the TACTIC  $\gamma$-ray    
telescope during 2004-2007. We do not find any evidence for the presence of a statistically significant VHE gamma-ray
signal  either in the overall data or   when the data are analysed  on a yearly spell basis   or a night to night basis  during the 2004, 2006 and 2007 observations. An upper limit of I($\geq$1.5 TeV)$\leq 1.18 \times10^{-12}$ $photons$ $cm^{-2}$ $s^{-1}$ (13.1$\%$ of the TACTIC detected Crab Nebula flux) has been placed at 3$\sigma$ confidence level on the integrated $\gamma$-ray flux. It may be noted here that we have not used the EBL absorption effect while deriving   the upper limit.  We conclude  that  the source was possibly  in a low TeV emission state (below the TACTIC sensitivity level) during this period. We have used    ASM data \cite{ASM} in the presented work  just to inspect the contemporary RXTE source light curves in the energy band of 2-10keV and  to compare with those obtained with the TACTIC observations.  In the earlier reported VHE detections, no evidence for the significant correlation between  X-ray and VHE $\gamma$ -ray   was found \cite{Horan02} as has been mentioned earlier.

In Figure \ref{fig:tev}, we show the available light curves of the source from 1999 to 2004   observed by the Whipple, CAT, CELESTE and GT-48 \cite{Horan02,Djannati02,david06,GT-48} groups.  It is interesting to point out here that the TACTIC 2004 results are in close agreement with those obtained by the CELESTE group \cite{david06}  around 100 GeV  threshold energy   during the same period. However, these results are in conflict with those obtained with the   GT-48 telescope\cite{GT-48} operating above 1 TeV  during the same period. At present we do not understand the reason for this conflict between these  TACTIC and GT-48  observations.

\begin{figure}[h]
\centering     
\includegraphics*[width=1.0\textwidth,angle=0,clip]{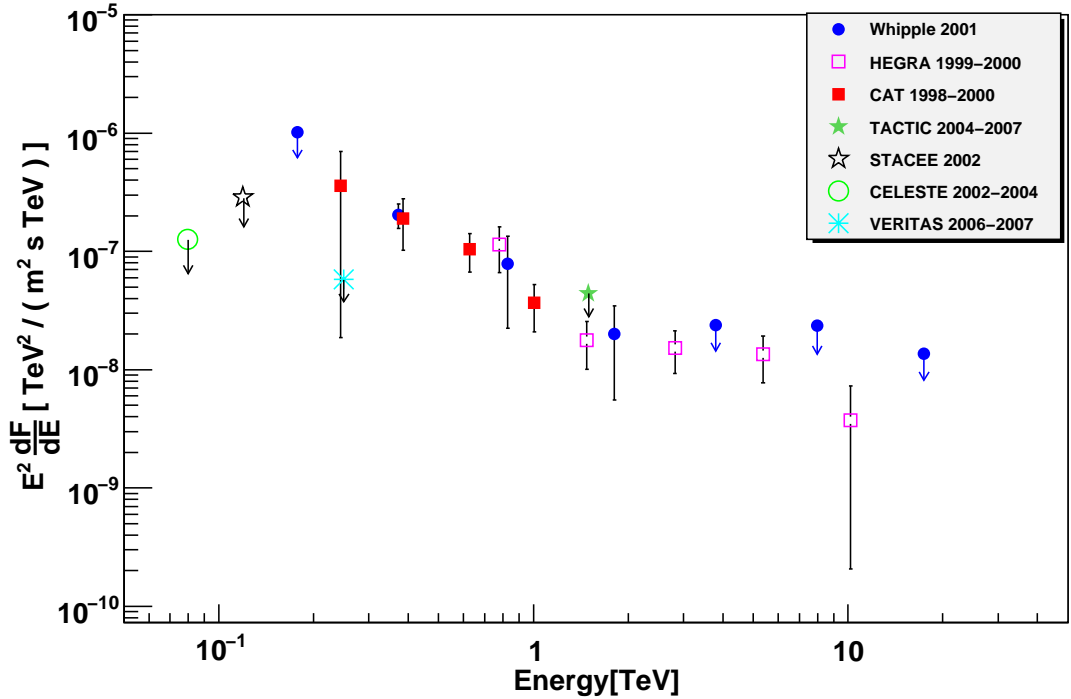}
\caption {TACTIC upper limit derived using 165.7 h of data has been shown along with results obtained by WHIPPLE\cite{Prtry02}, HEGRA\cite{Aharonian03}, CAT\cite{Djannati02} CELESTE\cite{david06}, STACEE\cite{stacee} and VERITAS\cite{kraw07} groups.}  \label{fig:comp}
\end{figure}

In Figure \ref{fig:comp} we have  compared the TACTIC  upper limit obtained during 2004-2007 with  the source   VHE spectra results  obtained during different epochs  by the   Whipple \cite{Horan02}, HEGRA \cite{Aharonian03} and CAT\cite{Djannati02} alongwith the upper limits provided by  the CELESTE\cite{david06}, STACEE\cite{stacee} and VERITAS\cite{kraw07} collaborations. Further, those groups who   have reported VHE $\gamma$-ray detections their  derived source spectra  are consistent with  the expected signature of absorption of TeV $\gamma$-rays by EBL \cite{Prtry02,Aharonian02,Aharonian03,Djannati02}.

\par
We  feel that  the long term  multi-wavelength  observations of the source  are required   to possibly  record more  flaring states of the source to understand it by using the  technologically advanced telescopes.  These  observations in the VHE regime would also possibly record a better statistics of VHE photons to derive the source related observed  and intrinsic spectra corresponding to various  high states. Indeed this would also help  in   understanding   the particle acceleration and $\gamma$-ray production mechanisms in the AGN jets, in addition to the  VHE photon horizon  and EBL related  quest. 

\par

\section {Acknowledgments} 
The authors would like to thank all the concerned colleagues of the  Astrophysical Sciences  Division  for their contributions towards the instrumentation and observation aspects of the TACTIC telescope. We  thank Professor  D. A. Smith and Professor V. V. Fidelis for providing useful information with respect to the CELESTE and GT-48 H1426+428 results respectively.  We   also thank the anonymous referees for their useful comments. 

\end{document}